# Teaching assistants' beliefs regarding example solutions in introductory physics

Shih-Yin Lin
*Department of Physics and Astronomy, University of Pittsburgh, Pittsburgh, PA 15213, USA*

Charles Henderson and William Mamudi
*Department of Physics and Mallinson Institute for Science Education, Western Michigan University, Kalamazoo, MI 49008, USA*

Chandralekha Singh
*Department of Physics and Astronomy, University of Pittsburgh, Pittsburgh, PA 15213, USA*

Edit Yerushalmi
*Department of Science Teaching, Weizmann Institute of Science, Rehovot, 76100 Israel*

As part of a larger study to understand instructors' considerations regarding the learning and teaching of problem solving in an introductory physics course, we investigated beliefs of first-year graduate teaching assistants (TAs) regarding the use of example solutions in introductory physics. In particular, we examine how the goal of promoting expert-like problem solving is manifested in the considerations of graduate TAs choices of example solutions. Twenty-four first-year graduate TAs were asked to discuss their goals for presenting example solutions to students. They were also provided with different example solutions and asked to discuss their preferences for prominent solution features. TAs' awareness, preferences and actual practices related to solution features were examined in light of recommendations from the literature for the modeling of expert-like problem solving approaches. The study concludes that the goal of helping students develop an expert-like problem solving approach underlies many TAs' considerations for the use of example solutions. TAs, however, do not notice and do not use many features described in the research literature as supportive of this goal. A possible explanation for this gap between their belief and practices is that these features conflict with another powerful set of values concerned with keeping students engaged, setting adequate standards, as well as pragmatic considerations such as time requirements and the assignment of grades.

## I.  INTRODUCTION

Helping students develop an expert-like problem solving approach is an important instructional goal for the introductory physics courses, valued both by physics instructors[1-3], as well as stake holders in academics and industry[4]. Indeed, instructional strategies have been developed and shown to help students improve their problem solving skills [5-15]. For example, the research comparing novice and expert problem solvers has identified expert-like approaches to problem solving involving elements such as initial problem analysis, planning ahead the solution, and ongoing evaluation of the progress made in the solution accompanied by refinement of the solver's understanding of the principles and concepts involved[16-20]. Researchers have proposed a prescriptive problem-solving strategy [9,13,21] as an instructional tool to explicate to some extent an expert-like approach to problem solving. This strategy includes three major components – (1) distinct problem analysis, (2) solution construction that makes explicit the plan, in particular intermediate goals and the principles used to figure them, and (3) checking the final answer. Research indicates that when instruction follows a cognitive apprenticeship approach (i.e. instructors explicitly model a prescriptive problem-solving strategy, require students to follow it, and coach them when doing so), students are likely to adopt more expert-like problem-solving approaches [5,7-12].

Although the effectiveness of these instructional strategies and relevant curricula materials has been documented in the research literature, such strategies and techniques are rarely put into common practice as intended[22]. Previous research on instructors' beliefs [18] that influence their choice of instructional strategies and material for introductory physics courses indicates that faculty tend to value, but not use aspects of these curricula. For example, although faculty generally held the learning goal of helping students develop expert-like problem-solving approaches and were aware of problem features that could support this goal (i.e. rich context, compound problems that are not broken into parts, etc.), they refrained from integrating these features in the problems they used during instruction because they believe these features conflict with a powerful set of concerns regarding clarity of presentation and minimizing student stress [2]. Such ambivalence could have served as a fruitful lever for reflective professional development processes. However, many faculty might not be open to engage in long-term professional development.

At many institutions, in addition to faculty members, teaching assistants (TAs) also play important instructional roles in the teaching of problem solving. For example, TAs are often the ones to lead recitations in which they present students with example solutions for physics problems, guide students in solving problems and assess students' solutions. A growing number of institutions require their teaching assistant to attend a training course and mentor them while they take their first steps as teachers [23-28]. Since teachers commonly construct their instructional beliefs and habits in their first years as teachers [29], the beginning of a TAs teaching career is likely a formative period that will influence later performance. Moreover, since most physics faculty have also been TAs, we expect it to be a formative stage where the beliefs of faculty develop. Thus, TA training courses provide a window of opportunity to significantly impact university physics instruction. To design effective professional development programs, it is important to know about TAs' beliefs when entering their job.

However, the research on TAs beliefs and practices in this regard is currently limited. This paper intends to fill in this gap by examining how the goal of promoting expert-like problem solving is manifested in the considerations of first-year TA choices of example solutions. (Please note, hereafter when referring to the TAs that we study, the term TA refers to the first-year TAs that we studied.) This instructional context was chosen for two reasons. First, it corresponds to one of the major responsibilities that TAs in many institutions are expected to fulfill (i.e. to guide students in problem solving and to present or demonstrate problem solutions in introductory physics courses). Second, in many of the aforementioned research-based instructional approaches [5-8,30] example solutions serve to explicate the problem solving processes of an expert. In this study, we will examine the extent to which TAs value modeling expert-like problem solving and how, if at all, they realize this goal through the design features of example solutions they would present to their students. For the purpose of our study, we use Reif's prescriptive problem solving method [9,13,21] as the standard against which we measure the features.

In particular, the study will focus on the following research questions:
(1) TAs goals and considerations for presenting example solutions:
   (1a) What are TAs' goals and considerations?
   (1b) Is developing an expert-like problem-solving approach a prominent goal?
(2) TAs awareness and attitude toward design features of example solutions that are aligned with the prescribed problem-solving model recommended in the literature
   (2a) Do TAs notice and value these design features?
   (2b) Do TAs use these features?
(3) Relationship between design features and goals:
   (3a) What design features of example solution do TAs perceive as supporting different goals/considerations?
   (3b) Where do conflicts exist (if any)?

## II. BACKGROUND AND LITERATURE REVIEW

This section elaborates the line of arguments made in the introduction above. We first describe instructional approaches recommended by the research literature to improve students' problem-solving approaches. We then discuss what is known about how example solutions can be structured to promote learning. Finally, we review what is known about faculty and TAs beliefs, in particular those related to the teaching of problem solving, and how these beliefs cohere with the recommendation made in the literature.

### A. Promoting expert-like problem solving approaches

There has been substantial work in the context of physics problem solving that has attempted to identify differences between experts (usually physics faculty and/or graduate students) and novices (usually introductory physics students) and to use these differences to help novices solve problems more like the experts. Several strong claims can be made from this body of research. The first is that compared to novices, experts typically employ a systematic approach when solving problems [13,31-33]. For example, experts typically devote considerable time at the outset for making simplifying assumptions that might help solve the problem and developing a qualitative re-description of the problem information in terms of physical quantities that are derived from an effective principle-based representation of the problem rather than representation that is focused on surface features [16,20,34]. This is true whether experts are presented with situations that are similar to problems they have already had practice with, or novel situations [35].

The second claim that can be made is that, based on this initial analysis of the problem, experts typically use the relevant information to plan the solution before executing it (e.g., grouping together solution steps into useful subproblems) while novice often follow a haphazard trial-and-error approach and perceive each solution step separately [20,32,36]. The third claim is that experts devote more time to assessing their solution process and their state of understanding (such as explicitly or implicitly asking themselves: What am I doing? Why am I doing that?). Experts are also much more likely than novices to evaluate their final answers [20,36,37].

Research indicates that when instructors explicitly model and encourage students to follow a set of problem solving strategies explicating expert-like behavior, students are likely to adapt a more expert-like problem-solving approach [5,7-12]. These strategies are also referred to in the literature [9] as prescriptive problem solving strategies. One should note that these prescriptive strategies reflect only some aspects of the actual problem solving process that an expert goes through and leaves out other aspects, such as back and force moves where the solver reconsider their approach to the solution. Several instructional techniques that have been used to promote expert-like problem solving involve "real" problems [5,6,14,15] that require a higher level of analysis and planning to encourage students to adopt more expert-like approach. They also involve modeling, coaching, and fading phases according to the cognitive apprenticeship framework [38]. Based on this framework, in addition to introducing problem-solving strategies that reflects the implicit problem-solving approaches used by experts, instructors should also demonstrate use of the strategies during class [9,13,21]. In particular, example problem solutions that instructors present to their students can be used to make explicit the prescriptive problem-solving strategies. Then, the students can be provided with opportunity to practice applying the strategies (e.g., by working with other students or with computers) under the assistance of the instructor. Scaffolding can be provided in one-on-one tutoring, by cooperative learning groups [7] and by computer-based tutoring [8]. As students develop more independence and expertise, the support from the instructor can be gradually reduced. In the following section, we briefly summarize the research related to how example solutions can be structured to promote learning.

## B. Structuring example solutions to promote learning

In this paper, we use the term "example solutions" broadly to refer to any problem solutions that students are exposed to during an introductory physics class. This includes solutions that the instructor works on the board during lecture and recitation, written solutions in the textbook or other similar material, and written solutions that the instructor distributes to students after students have submitted solutions to homework or test problems. Example solutions are used in nearly all introductory physics courses. They are described by a number of different names in the research literature, such as worked examples [39-42], worked-out examples [43-46], instructor solutions [47] and example problem solutions [1,3,48]. As mentioned earlier, research suggests that modeling the tacit problem-solving approaches of experts is an important instructional role of example problem solutions.

However, there is a difference between how good students and poor students study example solutions [46,49]. Students who "self-explain" the example solution more are more likely to benefit more from reading the example solution. Self-explanations refer to the content-relevant articulations (beyond what the text explicitly said) formulated by a student in order to make sense of an example solution [50,51]. This commonly involves filling in the gaps that correspond to the omissions in the solution and/or resolving the conflict between the students' mental model and the solution[51]. Chi [51] argued that high self-explainers are those who readily detect such conflicts while learning from an example solution. It is recommended that instructors provide students with prompts that encourage students to detect conflicts. Atkinson, Renkl, and Merrill [45] have shown that principle-based prompts are effective in inducing principle based self-explanations, one style of the self-explanation that successful learners use [43].

Research on the design of example solutions has shown that example solutions are more effective if multiple sources of information (e.g., diagram, text, and aural information) are integrated into a unified presentation [39,52-54] to avoid splitting students' attention across multiple non-integrated sources of information, which may cause cognitive overload on students [52,53]. In addition, structuring the examples to emphasize the important chunks of steps or subgoals (either by explicitly labeling them or simply isolating them visually) can guide students to discover the underlying deep structure of the solution and enhance learning [39,55-57]. With the aid of these structural cues, students are encouraged to explain to themselves why the steps are chunked together, which can promote the induction of generalization. It is found that, compared to students who learned from traditionally formatted examples solutions, students who learned from solutions in which subgoals were highlighted were more likely to transfer what they learned correctly into a new context [55]. Research also indicates that at the initial stages of skill acquisition, learning from example solutions is effective for improving problem solving performance compared to problem solving itself [39,58]. Because the cognitive overload is less when studying example solutions than actually solving problems, more short term memory capacity is available for students to extract useful strategies and to develop knowledge schemas [39,58,59]. At this stage, process-oriented solutions (solutions which present the rationale behind solution steps) are appropriate [41]. On the other hand, as learners acquire more expertise, process-oriented examples become less effective (or in some cases may even start to hamper learning) because the redundant information presented, which is hard to ignore, takes up unnecessarily large amounts of the limited working memory [41,60]. Product-oriented solutions (in which rationale is not included) are therefore more appropriate at the later stage of learning when the learners possess more prior knowledge.

Although educational researchers have provided insights into how the example solutions may be used to promote student learning, whether or not these research findings are incorporated into real practice by an instructor depends on the instructor's familiarity with the research findings as well as their beliefs about teaching and learning. In the following sections, we briefly summarize the research on instructors' (faculty and TAs) beliefs and practices.

### C. Faculty Beliefs and Practices

A number of researchers have investigated the general ideas of teaching and learning held by college instructors [61-67]. These studies generally found that instructors' views could be characterized on a continuum from teacher-centered, which emphasizes the transmission of knowledge, to student-centered, which emphasizes the construction of knowledge by the students. Instructors' beliefs about problem solving has also been found to be correlated with their general ideas about teaching and learning [66]. For example, instructors who conceived the meaning of the problem as obvious or unproblematic to students and thought of problem solving as an application of existing knowledge held more teacher-centered views of teaching. On the other hand, instructors who see that the meaning of the problem is not necessarily obvious to students and that problem solving involves making sense of the problem held more student-centered views [66].

Although instructors' ideas about teaching and learning influence their decision making, studies have found that due to conflicting factors or constraints, instructors' practices may not be consistent with the general ideas that they hold about teaching or learning [2,3,68]. A prior study [2] investigated the goals of 30 physics faculty related to their use of problems in their introductory physics course and found that "developing students' physics understanding" and "developing students' ability to plan and explore solutions paths" are two of the most mentioned learning goals that faculty expect problems to serve. In addition, faculty typically agreed with educational researchers about problem features that support these goals. For example, most faculty believe that developing students' ability to analyze problem situations and plan solution strategies requires having students cope with complex problems that require the use of these very same strategies. Examples of such problems advocated by curriculum developers are Context-Rich Problems [5], Experiment Problems [69], Real-world Problems [14], and Thinking Problems [15]. However, although faculty were aware of problem features that could support their goals, many of these instructors do not use these problem features and many even use features they believe hinder the goals. A strong reason for this misalignment comes from a powerful set of values concerning the need for clarity in the presentation and reducing the stress on students, especially during tests. Similar ambivalence was documented in two other studies [3,70] on how introductory physics faculty grade their student's problem solutions and how they construct example solutions for their students. In the first study [70], while all faculty reported telling students to show their reasoning in problem solutions (thus, making explicit how they analyze a problem solution and their plan to solve it), about half of faculty graded problem solutions in a way that would likely discourage students from showing this reasoning. Such grading practices provide insufficient feedback in order to promote in students expert-like problem-solving approaches. In the other study [3], it was found that although some faculty believe students would learn better from example solutions that contained more explanation of expert thought processes, they refrained from constructing these solutions for reasons including: (1) their inclination not to stifle student creativity in problem solving by providing example solutions that were too detailed, (2) their belief that an effective solution which conforms to the way an expert physicist would write a solution involves the shortest path to arrive at the result, (3) their concern that students may be frightened by problem solutions showing too many steps; and (4) their constraint of lack of time to construct such solutions.

### D. Teaching Assistants Beliefs and Practices

Several studies have been conducted to understand TAs' and Las' (Learning Assistants') beliefs and practices [23-25,71-74]. It is found that TAs' beliefs vary significantly regarding the role they think they should play in the classroom [73]. For example, in case studies of two TAs who teach tutorial-based recitations, one thinks of her major role as listening to student ideas and facilitating students' discussion [73]. When preparing for the recitation, she focuses on ways to explain or brainstorm interesting questions. On the other hand, another TA who thinks of his role as demonstrating expert reasoning prepares for the recitation by focusing on knowing the material thoroughly. It has also been shown that TAs' beliefs can affect their ability to carry out reformed curricula [24]. For example, a tutorial TA who doesn't believe that intuition from everyday experience can be a useful foundation for building physics knowledge may disregard students' common sense idea, which contradicts the intention of the tutorial design [24].

Since most TAs are graduate students a general understanding of graduate students' attitudes to problem solving in introductory physics provides another perspective on TAs' beliefs and practices. Mason and Singh [75] compare the attitudes and approaches to problem solving in the introductory physics courses by graduate students vs. introductory students. While close to 90% of the graduate students reported that they explicitly think about the underlying concepts when solving introductory physics problems, more than 30% of them perceived problem solving in introductory physics as essentially plug and chug. An examination of the graduate students' written explanations suggests that this view results from their relative expertise in solving introductory physics problems. Many of them can immediately realize the principle that should be used when solving a problem, therefore, perceiving the task as one that does not require much thought and reflection. Indeed, some graduate students noted that they do not need to reflect and learn from the problem solutions after problem solving in introductory physics because the problems are obvious to them and they feel that reflection is not needed.

Several institutions conduct TA and LA training programs [23-28]. A hallmark of successful training programs is matching the program to TA strengths and capabilities [27]. One of the goals of this study is to inform professional development providers to help them better support TAs to become more successful teachers.

### III. METHODOLOGY

#### A. Participants

Twenty-four first-year physics graduate students from the University of Pittsburgh were involved in this study. All 24 TAs were enrolled in a semester-long training course led by one of the authors (CS)[i], aimed at preparing them for their TA jobs. Most of the TAs in our study were simultaneously doing their TA job: 15 of them were working with the introductory algebra- or calculus-based physics course either leading recitations, lab sections, or being a grader, 2 of them were teaching astrophysics, and 1 of them was teaching an optical lab. The few TAs who were enrolled in the course and did not simultaneously have any teaching responsibilities were expecting to teach in the near future. In addition to teaching, these TAs also helped as tutors in the physics exploration center where introductory physics students could come for help in solving homework problems. Although many of the TAs were simultaneously doing their TA jobs, the study was conducted at the beginning of the semester when the TAs had just entered the graduate school. Therefore, it portrays the TAs' preliminary ideas when they had not have extensive experience in teaching introductory physics. Thirteen of the TAs involved in this study were international students, most of whom had their former undergraduate education in China or India.

#### B. Methodological consideration on the study of TA beliefs

Different methods have been used (and usually combined) by researchers to learn about instructors' conceptions and beliefs. For example, in addition to open-ended questions on a questionnaire or in an interview, researchers have combined open-ended questions with classroom observations [76] or descriptions of concrete hypothetical teaching situations [77,78]. Although instructors' teaching practices are influenced by their beliefs, simply observing the classroom practice may not reveal a complete picture because the actual practice may be a resolution of different beliefs (which may sometimes conflict each other) as well as situational constraints. Studies have also shown that similar teaching behaviors can be supported by different beliefs [72,73,79]. Thus it is difficult to infer beliefs based solely on observations of practice. On the other hand, the research literature also indicates that simply asking instructors (e.g., in an interview) about their conceptions is frequently not fruitful because conceptions can be implicitly held [80-83]. Interviews designed around concrete instructional settings have been shown to be successful in eliciting context-specific conceptions that may not otherwise be evident to the person who holds them [1,3,77,78].

This study introduces a methodology that is a variation of interviews designed around concrete instructional settings. We developed this methodology, called the Group Administered Interactive Questionnaire (GAIQ), to accomplish the following research goals [84]:

- Encourage TAs' introspection and articulation about their beliefs regarding the design of example solutions
- Triangulate findings regarding the above in various contexts, more and less concrete, more and less close to the actual decision-making
- Minimize distortion of data collection by researchers' personal bias (reliability)
- Compare results with pedagogies based on educational research

The GAIQ methodology represents a variation of interviews designed around concrete instructional settings that was used by two authors (Henderson and Yerushalmi) in a previous study of the considerations that shape faculty instructional choices regarding example solutions [1,3]. In the previous study, the data collection tool made use of semi-structured individual interviews that involved 2 types of questions: general questions about how and why the instructors might present solution to students, and concrete questions asking instructors to compare and make judgments about a set of artifacts that differed in features reflecting instructional approaches discussed in the research literature [1]. In the GAIQ methodology, the artifacts and questions asked to participants were similar to those used in the semi-structured individual interviews previously conducted. These artifacts and pre-determined questions serve to standardize data collection in order to collect reproducible data in both the interview and the GAIQ. However, in the GAIQ methodology, the individual interview was replaced with a written questionnaire (as shown in FIG. 5, FIG. 6 and FIG. 7) to respond to several concerns. First, from the practical perspective, individual interviews require significant time for both data collection and analysis. The GAIQ approach takes advantage of the TA training course to streamline data collection. Second, within semi-structured individual interviews, the interviewer interventions required to clarify respondents' answers may affect the reliability. In the GAIQ tool, clarification takes place via a sequence of worksheets, thus standardizing researchers' interventions. Finally, as the data collected in individual interviews is extremely rich, there is ambiguity in categorization of the data. The GAIQ approach incorporates elements intended to have the respondents clarify meaning (therefore achieving the validity of data suggested by Kyale [85]) by allowing the respondents to share and articulate their ideas in a group discussion as well as having them categorize some of the data themselves. The GAIQ methodology is described in more detail below and an explicit comparison between the interview and GAIQ approaches can be found elsewhere [84].

### C. The artifact comparison technique

The GAIQ methodology used an artifact comparison technique [1]. Respondents were asked to make judgments about instructional artifacts that were carefully designed to activate, in an imaginary classroom setting, the instructional decision-making that takes place in an authentic classroom. Through making and justifying instructional decisions, research subjects expose the beliefs and values that underlie these decisions in a way that is not possible through observational studies.

The artifacts were adopted from previous work conducted at the University of Minnesota [1]. They were three possible example solutions for a single problem selected to be one that could reasonably be given in most introductory physics courses. It was important that the problem be considered difficult enough by an instructor to require an average student to use an exploratory decision-making process as opposed to an algorithmic procedure. The problem is presented in FIG. 1. The example solutions (FIG. 2, FIG. 3 and FIG. 4) are designed to reflect various instructional styles in the actual physics classroom. None of the solutions were designed to be flawless. To frame the results discussed in this paper, we suggest the readers to take few minutes to look at the example solutions (FIG. 2, FIG. 3, and FIG. 4), reflect on how these solutions are similar or different to the solutions they use, and then try to articulate their reasons for favoring particular solution features.

The three example solutions differ from each other in important aspects. As shown in FIG. 2, example solution I is a "bare-bones" solution that leaves many of the steps to be filled in by the reader. This type of solution is typically found in textbook solution manuals. Although this kind of solution may make good sense to a problem solver who is more experienced and possesses more prior knowledge, it may be less effective for a beginning learner. Example solution II (FIG. 3) is more descriptive than solution I. It explicates many of the details of the solution process and represents another type of common actual instructor solution. Example solution III (FIG. 4), on the other hand, is designed to reflect a systematic decision making process characteristic of expert problem solvers along the prescriptive problem solving models suggested by Reif [9,13,21]. It begins with an overview discussing the problem goal and then relates the goal to the known information. The reasoning behind each step is explicated. Then, a separate "execution" section takes place to mathematically execute the plan. At the end, there is an assessment of the solution, which does not exist in example solutions I and II. There are other important differences between the solutions. For example, while example solutions I and II start with the knowns and invoke the conservation of mechanical energy principle to find the speed first, example solution III starts with the targeted variable and begin with the Newton's $2^{nd}$ law. Thus, example solution 3 reflects research findings that have shown expert problem solvers often begin with the problem goal and attempt to relate it to the known information. Although solution III represents many aspects of an expert-like problem solving process, it also misses some. For example, it is missing remarks like those found on the side of solution II that elaborates the rationale behind some steps.

The artifacts were validated as proper for an introductory physics course by a group of expert physics teachers/instructors [1,3]. The TAs at the University of Pittsburgh are generally expected to guide students in problem solving and to present or demonstrate problem solutions. Even if the TAs may not have extensive experience in teaching introductory physics, we do believes that the TAs would be familiar with the example solutions provided because these solutions were designed to reflect the actual practice in common physics classroom. In these regards, we expect the TAs to perceive the activities as relevant and reasonable. We made clear in the instruction that TAs should think as instructors when writing their own solutions and filling out the worksheets (Please refer to figures 5 and 6 for example.)

### D. Data collection procedure

Table 1 summarizes the GAIQ data collection sequence. In the pre-lesson stage, as part of their homework, TAs were asked to write a solution (to the problem presented in FIG. 1) that they would hand out to their students. TAs were then asked to respond to open-ended questions presented in FIG. 5 regarding how they think example solutions should be used in their instruction. The TAs were also provided with three example solutions for the problem (shown in FIG. 2, FIG. 3, and FIG. 4) and were asked to fill in a pre-discussion individual worksheet (FIG. 6) where they identified prominent design features of the solutions, ranked the solutions based on i) which solution has more of each feature and ii) their preference for including each design feature in solutions, and explained their reasons.

In the lesson stage the TAs interacted in small groups to share their ideas with peers, followed by a whole-class discussion. The lesson stage represented about 120 minutes of class time. In addition to the author who led the training course (CS), another author (SL) also sat in the classroom to take field notes.

Finally, in the post-lesson stage the TAs were provided with the opportunity to explain whether (and why) their preference changed by filling in a post-discussion worksheet (FIG. 7). On this post-discussion worksheet they were also asked to match the features they identified on the pre-discussion worksheet to a list of pre-defined features (presented in Table 2; TAs were only given the descriptions of each feature and not the information about clusters) corresponding to different aspects of the solution presentation. The list represents features identified in a pilot study with a similar group of first-year graduate students, who were enrolled in the same TA training course in the previous year and were asked to discuss the prominent features they observed in the same set of solutions artifacts on a worksheet similar to the pre-discussion worksheet used in our current study. The list of pre-defined features was created by coding the TA responses in the pilot study. With the help of this list of pre-defined features, the post-discussion worksheet allows the TAs in our current study to participate in the categorization of data, which promotes the validity of the study.

The features in Table 2 can be grouped into clusters to help in the interpretation of data. The first three clusters (C1 to C3) relate to the key stages in a prescriptive problem solving model described in the literature [13]. The final two clusters (C4 and C5) relate to communicating the solution to students. Each is briefly described in the following paragraphs.

The first cluster (C1) relates to the initial problem analysis as described by Reif [13]: "The purpose of the initial problem analysis is to bring the problem into a form facilitating its subsequent solution. To this end, one must first clearly specify the problem by describing the situation (with the aid of diagrams and useful symbols) and by summarizing the problem goals." (p. 27)

The second cluster (C2) relates to the search process that is the core of the solution construction stage described by Reif. The problem decomposition that takes place in this stage is materialized via feature 4 (explicit sub-problems are identified). Feature 6 (principles/concepts used are explicitly written) makes explicit the relations used in each sub-problem to eliminate unknowns. The rationale underlying the problem decomposition is described in feature 3 (providing a separate overview) and feature 5 (reasoning is explained in explicit words). A specific solution cannot demonstrate the recursive nature of the search process, yet feature 10 (providing alternative approach) reminds us that there are alternative solution paths. Finally, feature 12 (forward vs. backward solution) reflects possible directions of the search process.

The third cluster (C3) relates to checking the solution. For example, a symbolic solution (F13) allows one to check that the different stages in the solution are self-consistent. Performing a check of the final result by examining the unit or the limiting cases (F14) allows one to contemplate whether the final answer makes sense.

The fourth cluster (C4) and the fifth cluster (C5) are both related to the presentation of the solution. Features in the C4 clusters are all related to the "long/detailed" aspect of a solution. F7 (thorough derivation) and F9 (details without which the solution is still technically correct) represents two example components in a solution content that can lead to a long physical length (F8). On the other hand, the single feature in cluster C5 focuses whether the solution is presented in a clear and organized way (F11).

### E. Data analysis

The pre- and post-discussion worksheets as well as TAs' own solutions and their responses to the open-ended questions were collected for analysis.

To answer research question (1a), TAs' goals were identified by analyzing (i) their answers to the open-ended questions asking them to describe the situations and purposes for providing example solutions to students in a general context and (ii) their reasons for why they like or do not like the different solution features presented in the example solution artifacts in the concrete context. When analyzing the TAs' goals, open coding [86] was used to generate initial categories that were constantly compared to the new data and refined by the entire research team to arrive at a final set of categories. After developing coding categories, coding was done by one researcher (SL), with about 1/3 of the codes checked by other researchers who independently assigned the codes. Any disagreements were discussed by the researchers until full agreement was established. Approximately 7% of the coding was modified in this process.

To answer research question (1b), the numbers of TAs who identified the goal of developing an expert-like problem-solving approach along with the numbers of TAs who identified other goals in each context were counted.

To answer research questions (2a) and (2b), we focus on feature clusters C1 to C3 in Table 2, which relate to the goal of developing an expert-like problem-solving approach. For each cluster, we identified whether each TA (i) notices/values these features in the solution artifacts provided (question 2a) and (ii) makes use of these features in their own solutions (question 2b). For question (2a), to portray TAs preferences of features before being influenced by their peers, we focused on their answers from the pre worksheets, yet, as mentioned previously, we used TAs answers from the post worksheets (where they were asked to match the features they identified on the pre-discussion worksheet to a list of pre-defined features in Table 2) to assist us in clarifying the features and categorizing them. For each feature, TAs' preferences were determined by examining the reasons that the TAs wrote on the worksheet for why they like or do not like the feature and comparing whether the solution that each TA ranked as highest based on his/her preference for the feature matches the solution that he or she believed contain the most of this feature.

To answer research question (3a), we analyze all features that the TAs notice in the solution artifacts along with their descriptions of how they believe these features support or hinder different goals. As mentioned previously, the TAs were provided with 14 pre-defined features in the post-discussion stage and asked to identify whether the features they previously noticed can be matched to any of these pre-defined features. In addition to the 14 pre-defined features, we found that there were 3 additional features ("solution boxed", "meaning of symbols" and "in first-person narrative") that the TAs noticed. Because these additional features were only mentioned by 1 or 2 TAs, in the discussion of the results, we will focus only on the 14 pre-defined features.

To answer research question (3b), we take advantage of the result in research question (3a) and examine the extent to which features supporting different goals cohere with one another.

### F. Methodological limitations

We took a methodological approach of studying instructors' decision making in a simulated setting. However, this approach was not triangulated with TAs' actual practice. Although the TAs were explicitly instructed to perform the task as an instructor while writing their own solutions and filling out the worksheets, there is no direct information about the extent to which TAs actual practice resemble the perceived practice observed in the TA training course. An observation of TAs actual practice can be carried out in the future to shed light on this issue.

In addition, the GAIQ methodology, which was intended to encourage negotiation of meaning in a survey context, only partially achieves this goal. For example, we found from our study that the TAs written responses were concise relative to spoken ones. Measures should be taken to encourage the future respondents to explicate their opinions on the worksheets in more depth, such as enlarging the spaces provided in the worksheets, or providing an example of elaborate response to a hypothetical feature.

## IV. RESULTS

### A. TAs goals and considerations for presenting example solutions

#### 1. Question (1a): What are their goals and considerations?

In order to discriminate between strongly held beliefs and fragmented or even conflicting ones, we asked TAs about their goals (and/or considerations) in two contexts: (i) a general and open ended context (e.g., TAs were asked "What is the purpose you see for providing solved examples in different situations?", etc.); (ii) a concrete context simulating to some extent the decision making that takes place in actual practice (i.e. TAs were asked to describe why they would or would not include particular design features in the example solutions they present to students). We expected the goals that the TAs express might differ between the general and the concrete contexts. While in a general context instructors might express commitment towards possibly conflicting goals, the actual practice requires them to resolve such conflicts, exposing their priorities, some of which the instructor might not be conscious of (i.e. depth vs. time concerns, etc.). In addition, the design of the solution artifacts used in the concrete context aimed to elicit possibly deeply held ideas regarding "developing an expert-like problem-solving approach". Such ideas are likely to be uncovered in the general, open ended context only if they are on top of the interviewees' minds. We hypothesized that if such ideas are implicitly held, the artifact comparison technique may allow the TAs to articulate these ideas since the solution artifacts differed from each other in the extent to which they reflect a systematic decision making process characteristic of expert problem solvers.

As discussed above, the goal categories were developed via an emergent coding procedure. The goals are summarized in Table 3 and then discussed in more detail in the following paragraphs. Note that some of the goals were only present in one of the two contexts (either general or concrete). The first two goals (G1 and G2) in Table 3 relate to expected performance of the students. Thus, we will refer to them as learning goals. The additional goals (G3 to G7) represent other considerations that the TAs expressed. These include considerations intended to support students' learning process (G3 to G5) and pragmatic concerns regarding time and student grades (G6 and G7).

*Learning goals*
G1 – physics understanding: An example solution should help students construct content specific physics understanding. This goal is expressed in both the general and concrete contexts, by 23 and 18 TAs, respectively. For example, the example solution should help to clarify concepts (e.g., "Examples make abstract concepts more concrete by showing how the concept is applied." (TA4, general context)), relate concepts to problem solving (e.g., "It is most important to solve problems during the lecture, so that the students can relate the problems to the concepts they are being taught" (TA3, general context)), expand the breadth of possible approaches to solve a problem (e.g., "knowing how to do things several ways helps" (TA15, concrete context)), and refine understanding (e.g., "They [students] should be given solved problems after the homework is due but before the test so that they can prepare for the test and better understand where they may have gone wrong in their approach" (TA21, general context)). The TAs also believed that a good solution should help students focus on concepts rather than equations (e.g. "students should learn and memorize concept, not just equations so they can apply them" (TA2, concrete context)). All 18 TAs who mentioned this goal in the concrete context also mentioned it in the general context.

G2- PS approach: An example solution should help students develop an expert-like problem-solving approach. As we hypothesized above, the artifact comparison technique helped the TAs to recall and articulate this goal: it was expressed by few TAs in the general context (by 5 TAs) and by many in the concrete context (by 17 TAs). For example, in the general context, a TA expressed that "I think seeing problems in lecture provides students with the general idea of how to approach problems" (TA15, general context). When discussing the valued design features in an example solution (concrete context), the TAs believed there are several tools that should be included in the example solutions because these tools facilitate desired thinking processes that are characteristic of expert-like problem solving. For example, a diagram "allows you to visualize [the problem]" (TA1, concrete context), and doing a unit check at the end "allows students to evaluate their final answer-does it make sense" (TA5, concrete context). In total, this goal is expressed by 18 unique TAs (in either the general context, the concrete context, or both.)

*Other considerations*
G3 – emotional engagement: This goal is expressed by 9 unique TAs in total. It is expressed to the same extent in both the general context (by 5 TAs) and concrete context (by 6 TAs). However the meaning in these two contexts differs slightly. In the general context it was expresses in terms of students' confidence: Example solutions can be used to motivate students or to prevent student frustration. For example, "[After students seeing an example solution to a similar problem in the lecture,] when they try the [homework or exam] problem on their own, they can reference the different problem solving strategies. In this way, they aren't just starting blindly and they won't get as frustrated." (TA5, general context); "It is also useful to see the abstract concepts used in a practical way in order to give the students motivation to learn the material" (TA6, general context). In the concrete context it was expressed in terms of maintaining students' interest. For example, some TAs explained that they don't like to have a detailed solution "[because students] won't have patience to finish it" (TA16, concrete context). Other explained that the solution should "easily explain concepts without scary math" (TA6, concrete context).

G4- cognitive engagement: This goal is also expressed to a much lesser extent in the general context (by 6 TAs) as compare to the concrete context (by 21 TAs), and the meaning in these two contexts differs slightly. As described in the general context, example solutions should be presented at a proper time to engage students in cognitive processing. For example, "since everybody has to attend test and they tried very hard to solve that problem, so this [after the test] is a best time to explain some common mistakes" (TA10, general context); "Solutions can be useful only when the problems have been considered. If we give them the solution during lecture or before homework, they will copy it and it's totally useless" (TA22, general context). As described in the concrete context, example solutions should be communicated in a manner that allows students to follow it. Thus, solutions should be "easy to understand" (TA17, concrete context) and avoid the situation where "someone who is lost could not follow this" (TA6, concrete context). We believe that observing concrete artifacts that are less and more easy to follow raised TAs awareness to this aspect. In total, this goal is expressed by 22 unique TAs (in either the general context, the concrete context, or both.)

G5 – setting standards: This goal is expressed only in the concrete context. It is expressed by 12 TAs. There are some features that the TAs like because they are considered as the standard for an adequate solution. For example, the solution should include the solution process (e.g., "always tell the students to show work" (TA2)), be efficient (e.g. "physics is straight, it should be solved in the most simple way" (TA22)), and orderly presented.

G6 – saving time: This goal is expressed only in the concrete context. It is expressed by 5 TAs. Some TAs like a concise solution because a short solution "saves time" (e.g., TA3, TA 20). One TA explicitly points out that "[a solution with fewer steps] can save time in exam", which reveals his concern about saving students' time in a situation in which time is essential. Although the other TAs didn't specify whose time is saved (the TAs', or the students') on their worksheets, in the whole class discussion TAs mostly expresses concern about their own time as they were busy with their own graduate course work.

G7 – preventing mistakes: This goal is expressed only in the concrete context. It is expressed by 2 TAs. These TAs feel that a concise solution can lower the possibility of making mistakes (e.g. "less steps, less mistake" (TA19), "more simple, less mistake" (TA22))
. If students are presented with concise solutions, they may learn to refrain from length to avoid lowering their grades

*2. Question (1b): Is developing an expert-like problem-solving approach a prominent goal?*

The percentage of TAs who mentioned each goal/consideration in the general and concrete contexts, respectively, are shown in FIG. 8. Developing an expert-like problem-solving approach is, indeed, one of the prominent goals expressed by the TAs. FIG. 8 indicates that the goals/considerations the TAs mentioned differed between contexts. In the general context (responses to open-ended questions) the TAs discussed example solutions mainly as tools to help students construct content-specific physics knowledge. Only 21 percent of the TAs mentioned developing an expert-like problem solving approach when asked to generally describe their goals for example solutions. On the other hand, when considering the concrete instructional artifacts, approximately equal numbers of TAs considered example solutions as a means to help students develop an expert-like problem-solving approach (71 percent, corresponding to 17 out of 24 TAs) and as a means to help students construct content-specific physics knowledge (75 percent, corresponding to 18 out of 24TAs). The prominence of the goal of helping students developing an expert-like problem-solving approach in the concrete context can likely be attributed to the design of the solution artifacts aimed to elicit possibly implicitly held ideas related to the development of expert-like problem-solving approaches. These results suggest that, while TAs value this goal when they see it materialize in specific design features, they may not be explicitly conscious that they value this goal.

Given this result, we proceed now to examine how this goal of helping students develop an expert-like problem-solving approach materializes in practice. In the following sections, we first investigate whether the TAs notice, value, and make use of features that the literature perceives as supporting this goal. Then, we examine how other goals/considerations interfere, if at all, in materializing this goal.

### B. TAs awareness and attitude toward design features that are aligned with the prescribed problem-solving model recommended in the literature

To examine TAs awareness and attitude toward features that the literature identifies as supportive of the goal of helping students develop an expert-like problem-solving approach, we focus on the clusters grouped in Table 2 that relate to key stages in Reif's prescribed problem-solving model [9,13,21]: C1- initial problem analysis; C2- solution construction;

C3- checking of solution. The results to questions (2a) and (2b) will be discussed three times: once for each of the three problem solving clusters.

### 1. Features Related to Initial Problem Analysis (C1)

(1) *Question (2a – C1): Do TAs notice and value design features related to C1?*

Providing a schematic visualization of the problem (F1) and providing a list of knowns/unknowns (F2) are the features that relate to the explication of the initial problem analysis stage in a prescribed problem solving model. F1 is one of the most mentioned features (13 out of 24 TAs). F2 was mentioned by 9 TAs (the median for all features). As FIG. 9 shows, these features were valued by almost all TAs who mentioned them. Only one TA expressed that he didn't like to provide a list of knowns/unknowns because it encourages students to solve problem via mindless plug and chug. Other TAs valued the list of knowns/unknowns because it "gives an idea of what you have and what you need."

Although all TAs who noticed F1 (visualization) valued it, different TAs had different ideas about the preferred visualization shown in FIG. 10. Nine out of thirteen TAs distinguished between the qualities of diagrams, with 6 of them preferring a detailed drawing as presented in solution 3. Most of the TAs did not articulate why the detailed diagram was better than the others. TAs who chose the less detailed diagrams in solution 1 and/or 2 explained, for example, that they didn't like diagram 3 because "complicated diagrams can be confusing" (for example, the arrows in diagram 3 could be confusing to the students because they are used to represent both acceleration and velocity).

(2) *Question (2b– C1) Do TAs use design features related to C1?*

Examination of TAs' own solutions (which 23 TAs provided) indicates that features related to initial problem analysis are surprisingly more prominent in TAs actual practice than in their account of liked features. All TA solutions included a diagram and half included a list of knowns. A possible explanation resides in the fact that all three solution artifacts included a drawing, which made this a less noticeable feature within the artifact comparison technique.

### 2. Features Related to Solution Construction (C2)

(1) *Question (2a – C2) Do TAs notice and value design features related to C2?*

Six features (F3, F4, F5, F6, F10, F12, see Table 2 in the Methodology Section) relate to the solution construction stage in a prescribed problem solving model. Based on Reif's [13] representation of problem solving as a decision making process, the major choices a person makes in a solution process involve defining sub-problems: intermediate variables and principles to find them. These two aspects are evident in features F4 (Explicit sub-problems are identified), and F6 (Principles/concepts used are explicitly written). We grouped them under "Choices made" group. While F4 and F6 describe the major choices one makes, F3 (Providing a "separate" overview) and F5 (Reasoning is explained in explicit words) describe the solver's reasoning underlying these choices. We grouped them under "Reasons for choices (additional explanations)" group. We note that this reasoning is guided by the solver's general perception of the framework within which choices are made (e.g., as a process that involves choosing between alternatives, or arriving at identified goal in a backward manner) represented in F10 (Providing alternative approach) and F12 (Forward vs. backward solution).

FIG. 11 shows that features related to *reasons for choices* (F3 and F5) were the most noticed ones, but even those were noticed by less than half of the TAs (9 and 11 TAs, respectively). More than half of the TAs who noticed feature F3 thought that it was best represented in solution 3 (FIG. 4). Indeed solution 3 describes a complete overview of how the problem should be broken into sub-problems and explains the principles applicable in each of the sub-problems at the very beginning. As for feature F5, more than half of the TAs who noticed this feature thought that it was best represented in Solution 2, shown in FIG. 3. This solution identifies the goal of each sub-problem and provides justification for the principles separately as the progress of the solution. Although most TAs did not explicate why one presentation of F5 is better than the other in the worksheets, in the whole-class discussion several TAs raised their concerns that students may not have the patience to read the whole chunk of text at the beginning of solution 3. Students may simply ignore all the explanations in the first part and jump directly into the second part with equations. Reasoning that is presented beside the equations, as in solution 2, makes it easier to reference and students are more likely to process the information better.

As shown in FIG. 11, we find that F3 and F5 were valued by most TAs who noticed them. The TAs believed that these two features play an important role in example solutions because they make the solution process clear and make the solution easier to follow. The TAs also believed that these features help students understand the internal thinking process that the instructor went through when solving the problem and facilitates better transfer to other problems. Except for minor concerns, such as "overdoing the motivations can lead to undesired chunks of text", which was the major reason why a few of the TAs expressed a conflicted preference, these features were generally valued by TAs.

Features 4 and 6, which explicate the *choices made*, were noticed by only a few TAs (2 and 5 TAs, respectively), although they were valued by all TAs who noticed them. One TA explained that "I enjoy this feature [F4] because it helps set up a logical progression of the problem"; other TAs explained their preference towards F6 in that "the concepts may be more important than the answer" or "if we can use less math, I think we should do that, so students focus on physics".

Regarding the *framework within which choices are made*, 4 of the 5 TAs who noticed F10 (providing alternative approach) liked this feature, explaining, for example, that "this [feature] demonstrates how to develop an expert knowledge structure and how it makes the problem much simpler." One TA was conflicted about this feature, as presenting an alternative approach "could possibly confuse students." As for F12 (backward vs. forward solution), only one TA noticed it as an important consideration in the design of a solution.

(2) *Question (2b– C2) Do TAs use design features related to C2?*

Examination of TAs' own solutions indicates a discrepancy between their self-reported preferences and their actual practice. Although F3 and F5 were valued by 7 and 10 TAs, respectively, only 3 out of 23 TAs provided some outline of the sub-problems (F3) either at the very beginning or along the solution progression, and only 6 of the 23 TAs provided any justification for the principle(s) used (F5). None of the TAs presented a solution in which the goals for each sub-problem were clearly stated (F4). None of them provided an alternative approach in their own solutions (F10), either. On the other hand, slightly more TAs implemented F6 than the number of TAs who reported to notice and value this feature. The concepts of both "conservation of energy (COE)" and "Newton's $2^{nd}$ Law (NSL)" were explicitly written in words or the basic mathematical forms by 7 TAs. It is likely that for some of these TAs, explicitly writing the principles used (F6) is their natural practices while solving a problem but they didn't necessarily think about the importance of this feature in terms of its instructional implication.

Out of the 6 features in cluster C2, the greatest difference between TAs' self-reported notion and their actual practice was observed in F12 (backward solution), an important feature suggested in the educational literature. Research indicates that one difference between experts and novices is experts (teachers) commonly regard introductory physics problems as exercises while they are actually problems for novices (students). As a result, experts may present problem solutions in a forward manner, reflecting their knowledge of the problem solution in an algorithmic way. Yet, to explicate the decision making process of an expert when solving a real problem, one has to present the solution in a backward manner. The only TA who mentioned this feature, however, presented his solution in a forward manner. On the other hand, there were 8 TAs who originally presented a backward solution, even though they did not mention F12 in the worksheets. This suggests that many of the TAs do not recognize the backward and forward solutions as distinct features.

### 3. Features Related to Checking of Solution (C3)

(1) *Question (2a – C3) Do TAs notice and value design features related to C3?*

Feature 13 (symbolic solution) and feature 14 (providing a check of the final result) are the features which are related to the $3^{rd}$ stage in the prescriptive problem-solving model: checking of solution. Having a symbolic solution (F13) is important in the teaching of physics problem solving because many students tend to plug in the numbers at the beginning of their solutions. Only 2 TAs noticed this feature. The fact that most TAs didn't notice this feature may be due to the fact that solving problems symbolically has become a natural practice for them and because that this feature was present in all 3 solution artifacts provided. On the other hand, the fact that many more TAs (13 TAs) noticed F1 (providing a schematic visualization) than F13 even though both features can be found in all 3 solutions suggests that F13 is a deeper feature that may require a deep familiarity with the teaching of physics problem solving in order to be able to notice it.

As for feature 14, we expected this feature to stand out in the artifact comparison technique since only 1 of the 3 solutions included it. However, as shown in FIG. 12, only 4 TAs noticed this feature.

(2) *Question (2b– C3) Do TAs use design features related to C3?*

Similar to cluster C2, TAs' actual practices don't match well with their awareness and preferences for features in cluster C3. Although only 2 TAs mentioned feature F13, examination of TAs' own solutions show that all TAs solutions were symbolic. On the other hand, although F14 was valued by all the TAs who noticed it, only one TA performed an answer check in the solution he prepared for the introductory students. The findings suggest that TAs did not make use of the symbolic nature of their solutions to check the final result. It is likely that the importance of F14 was underrated or ignored by most of the TAs.

## C. Relationship between design features and goals

In the previous section, we examined the extent to which TAs value/use features that the literature suggests as supportive to the goal of helping students develop expert-like problem-solving approaches. Here, we look at the extent to which the TAs themselves connect these features to that goal. Moreover, we investigate how other goals and considerations interfere with the use of features that help students to develop an expert-like problem-solving approach. In order to do so, we first look at question (3a), examining all the features and their relationship to the goals that the TAs expressed.

### 1. Question 3a: What design features of example solutions do TAs perceive as supporting different goals/considerations?

To get a somewhat more global picture, the features' (grouped into clusters as described in Table 2) relation to the goals is portrayed in figure 13. The x-axis states the various goals/considerations that the TAs mentioned. For each of these goals, the number of TAs who perceive different features as supportive or contradictive to the different goals are plotted. The height of each bar indicates the number of TAs who noticed at least one feature in that particular cluster and believed that the feature(s) support (represented as positive) or contradict (represented as negative) that goal. For example, if a TA noticed F3 and F10 – both from the C2 cluster – and valued them because s/he believed these features can help students develop expert-like problem solving approach (G2), this TA is represented as one single unit in the gray positive chunk of the G2 bar. On the other hand, if another TA noticed F10 and dislike this feature because s/he believed that this feature would make the solution confusing, therefore hindering goal G4, this TA is represented as one single unit in the gray negative chunk of the G4 bar. It is possible that a single TA may consider features from different clusters as supportive to the same goal/consideration. Therefore, for each goal, the maximum value on the vertical axis could potentially sum up to 120 (24 TAs times 5 clusters).

As one might expect, the analysis (C1), construction (C2) and checking (C3) clusters - clusters that are aligned with the prescribed problem solving model by Reif - are indeed those that TAs recognize as supporting the goal of helping students develop an expert-like problem-solving approach (the G2 shown in FIG. 13). Moreover, these three clusters are also valued by the TAs because they support some other goals/considerations. For example, the *solution construction* (C2) cluster is the most prominent cluster in *helping students construct content specific physics understanding* (G1- physics understanding) and is also prominent in promoting "*cognitive engagement*" (G4).

In addition to clusters C1, C2, and C3, another prominent feature shown in FIG. 13 is the *extended details* cluster (C4), which is considered as disadvantageous in regard to many considerations (such as G3 – *emotional engagement*, G6 - *saving time* and G7 – *preventing mistake*), and in some cases both positive and negative (such as G5- *setting standards* and G4-*cognitive engagement*). For example, although a detailed solution may make it easier for students to follow, it could also work in the opposite way and make the students lose the thread more easily.

We now turn to examine whether the above portrayal of goals and practices challenges the materialization of the goal of helping students develop an expert-like problem-solving approach.

## 2. Question 3b: Where do conflicts exist (if any)?

The above portrayal of relationships between the features and goals, in particular, the positive occurrence of cluster C2 (*solution construction*) and the frequent negative occurrence of cluster C4 (*extended details*), suggests a possible conflict between design features supporting the development of an expert-like problem-solving approach (G2) and design features supporting other goals/considerations. While C2 is perceived as one of the prominent clusters supporting goal G2, it usually requires a longer length and more details, which is represented by the cluster of *extended details* (C4). However, C4 is considered as disadvantageous in regard to many considerations. Such finding suggests a possible challenge to materialize the goal of developing an expert-like problem-solving approach and other goals in a coherent manner.

In order to examine whether such a conflict actually exists within individual TAs and to get more insight into the nature of the conflict, TAs' preferences for cluster C2 are compared to their preferences for cluster C4. In particular, we focus on features F3 (*providing a "separate" overview*) and F5 (*reasoning is explained in explicit words*) in the *solution construction* - C2 cluster and features F7 (*thorough derivation*) and F8 (*long physical length*) in the C4 cluster. F3 and F5 were chosen because they are the features in cluster C2 that most require extended details. F9 (details without which the solution is still technically correct) was excluded from the features in C4 (leaving F7 and F8) because F9 is already stated in a negative manner.

In particular, to gain insight into how TAs resolve the conflict we: (i) identify how TAs holding conflicting values between different features (even though they may not be aware of it) resolve this conflict in their actual practice, (ii) identify the direction in which TAs shift their preferences for a single feature between the pre and the post worksheets, (iii) identify TAs resolution for conflicting preferences and concerns regarding a single feature that they were aware of and raised in either one of the worksheets itself. In total there were 13 unique TAs who fell into one or more of these categories (with 9, 3, and 6 TAs falling into categories (i), (ii), and (iii), respectively.) Each category is discussed in detail in the followings.

For (i), out of the 14 TAs who mentioned at least one feature from each of these two groups (F3/F5 vs. F7/F8), 9 of them were found to hold a conflicting value: while they valued F3 (*reasoning*) and/or F5 (*overview*) in the C2 cluster, they dislike the feature(s) F7 (*derivation*) or F8 (*long*) in the C4 cluster. Examining their own solutions indicates that except for one TA's solution which is detailed, most of these TAs' solutions were concise, and there is little, if any, F3 (*reasoning*) or F5 (*overview*) found in them. Thus, when TAs were not aware of the conflict, in most cases their resolution gave up features supportive of expert like problem solving.

However, when TAs were aware of the conflict, the resolution may have taken a different direction. For (ii), it is observed that after the group discussion, three TAs indeed re-considered their former preferences regarding the design of problem solutions and explicitly changed their preferences from concise solution to thorough presentation after the group discussion. Reasons why they initially preferred a concise solution include "less exhaustive, more efficient" (TA 24) and "use the best solution with least steps" (TA 23). After the group discussion, they focused on a different goal and explained that they preferred a longer solution because "appropriate physical length will help student follow the steps" (TA 18) or that "if it's too concise, people may be confused" (TA 23).

For (iii), six TAs were found to explicitly express a conflict about the use of a design feature. The different concerns raised signify that these TAs were aware of the challenge in materializing the goals coherently. For example, although one TA (TA 14) consistently preferred a concise solution (the opposite of F8) in the pre and post worksheets, he raised his concern about the disadvantage of this feature in the post-discussion worksheet, noting that "Solution 1 is short and sweet, hard to understand for a layman though". Another TA (TA 3) readily raised the similar concern regarding a concise solution (in the context of the opposite of F9) in the pre-discussion worksheet, noting that "[conciseness] saves time, but could cause confusion." Three TAs (TA 7, TA 15, and TA 13) expressed that it is necessary to find a middle ground between conciseness and explanation.

As we became familiar with the data, we realized that TAs' former educational backgrounds may play a role in their preferences related to clusters C2 (*construction*) and C4 (*details*).

Table 4 shows that before the peer discussion, Non-American TAs (N=13), most of whom had their secondary and prior post-secondary education in China or India, were more likely than American-educated TAs (N=11) to dislike F7 and/or F8 even though both groups tended to like F5 and/or F3.

TAs from foreign countries may have different expectations about what an introductory physics student is able or expected to do. As one TA who was formerly educated in China explicitly pointed out after the activity: "TA solution should be clearer than just a few key steps. That's what I really learned. In the class, all of the native students [TAs] tended to avoid using a simple key step solution. That's surprising because in my own country I have only seen such solutions. I used to avoid using many words explaining what is going on and why we have to apply these theorems, because that's the situation in my own country, where students have to think all by themselves" (TA 12). This statement echoes the result discussed previously in which some TAs re-considered their former preferences regarding the design features in the example solutions after the discussion with peers.

## V. CONCLUSIONS

### A. TAs' Goals related to the use of example solution

In this study, we find that helping students develop an expert-like problem-solving approach – a goal that is aligned with recommendations from educational research - is indeed a prominent learning goal that most (75%) of the TAs expressed when contemplating the use of example solutions. This result is aligned with prior research, which found that "helping students develop an expert-like problem-solving approach" is one of the most mentioned goals underlying physics faculty decision-making related to different problem or solution features [2,3]. We also find that this goal is mentioned by the TAs more frequently when looking at concrete artifacts than when asked general questions about their goals. This result suggests that this goal may be implicitly held by many of the TAs.

In addition to the goal of helping students develop an expert-like problem-solving approach, the TAs also expressed several other goals/considerations that are related to student learning as well as pragmatic issues about time and grades. The most prominent of these goals/considerations are helping students construct content specific physics understanding (G1), and designing the experience of studying an example solution in a manner that will be cognitively engaging (G4), expressed by 96% and 92% of the TAs, respectively. Using example solutions to send a message regarding expected standards (G5) was expressed by half of the TAs. Other considerations were much less prominent: emotional engagement (G3, expressed by 38% of the TAs), saving time (G6, expressed by 21 % of the TAs) and preventing mistakes (G7, expressed by 8% of the TAs).

B. Do TAs notice, value and use design features aligned with the prescribed problem-solving model that has been shown to help students develop more expert-like problem-solving approaches?

Although helping students develop an expert-like problem-solving approach is a prominent goal that the TAs expressed, many design features which are aligned with the prescribed problem-solving model were noticed by only few TAs. The most prominent feature - schematic visualization of the problem (F1) in the C1 cluster *(initial problem analysis)* was noticed by 54% of the TAs. In the *solution construction* cluster (C2), features which explicate the solver's reasoning underlying their solution choices (F3 – providing a separate overview and F5 – reasoning is explained in explicit words) were noticed by about 40% the TAs. However, other features in the C2 cluster, which explicate the choice of sub-problems (F4 - Explicit sub-problems are identified and F6 - Principles/concepts used are explicitly written) and the framework within which choices were made (F10 – providing alternative approach and F12 – backward solution) were noticed by very few TAs. Features in the C3 cluster (*checking of solution*) were also noticed by very few TAs. The fact that the TAs didn't notice many of the features in these clusters indicates that the TAs are not aware of the types of scaffolding recommended in the research literature as ways to help students extract an expert-like problem-solving approach from example solutions

In addition, the TAs' self-reported preferences didn't match well with the solutions that they wrote on their own before seeing the three example solution artifacts. Although features in all three clusters that are aligned with the key stages in a prescriptive problem-solving model were in general valued by the TAs, only features related to the initial problem analysis (especially schematic visualization - F1) and the symbolic presentation (F13 in C3) were typically found in their own solutions. The majority of the TA solutions contained little or no reasoning to explicate their underlying thought processes behind solution choices. No TA presented a solution in which the goals for each sub-problem were clearly stated. No TA provided an alternative approach. An answer check was found in only 1 TA's solution.

C. Materialization of the various goals – how do other goals and practices interfere with materializing the goal of developing an expert-like problem solving approach?

As expected, feature clusters C1 (*analysis*), C2 (*solution construction*) and C3 (*checking*) – clusters that are relevant to the explication of expert-like problem solving from a theoretical point of view – are the prominent features that the TAs recognized as supporting the goal of helping students develop an expert-like problem-solving approach (G2). These features were also recognized by the TAs as supportive of some other goals (such as helping students construct content specific physics understanding (G1) or cognitive engagement (G4)). However, an implicit conflict between design features supporting the development of an expert-like problem-solving approach and design features supporting other goals was observed in this study. In particular, feature cluster C2 (*solution construction*), a prominent cluster that TAs believe to support the goal of helping students develop an expert-like problem solving approach, usually requires a longer and more detailed solution, which is represented by feature cluster C4. However, feature cluster C4 is considered disadvantageous by many TAs because of its contradiction to goals/considerations such as G3 (*emotional engagement*), G4 (*cognitive engagement*), G5 (*setting standa*rds), G6 (*saving time*) and G7 (*preventing mistakes*). In most cases, when TAs were not aware of the conflict they resolved it in favor of a brief solution, giving up features supportive of expert-like problem solving. However, when they were aware of the conflict their resolution was likely to differ, either towards a "midway" or towards longer solutions that are supportive of expert-like problem solving.

Moreover, in this study, we find that TAs' conceptions of the goals/considerations and the preferences for corresponding features are influenced by their former education. On average, TAs with foreign background were more likely to value product-oriented solutions (in which rationale is not included) as compared to the American TAs.

In summary, in this study we find a gap between goals and practice. For most TAs, the goal of helping students develop an expert-like problem-solving approach underlies their choices of example solutions in introductory physics. Yet, when choosing specific example features, they do not notice, and do not use many features described in the research literature as supportive of the goal of helping students develop an expert-like problem-solving approach. A likely explanation for this gap is that the very same features helping to explicate an expert-like problem-solving approach along a prescribed problem-solving model also require extended details. These details are perceived by TAs as detrimental to their other considerations of engaging student cognitively and emotionally, setting adequate standards, and the pragmatic concerns about time and grades.

## VI. IMPLICATIONS

A recent study suggests that a way to improve TA professional development is to "find and build upon productive elements in their beliefs" (Ref. [71], p. 1). Our current study sheds light on TAs' beliefs at the beginning of graduate school, and provides comparison to practices and considerations suggested in educational research that help students develop more expert-like problem-solving approaches. Accordingly, the results of this study can inform strategies to support TAs in improving their instructional practices. From a professional development point of view, the TAs' view that example solutions should help students develop more expert-like problem-solving approaches is a productive starting point. Although the TAs did not necessarily notice all design features that can help with this goal, it is likely that the use of specific artifacts to elicit TAs' initial ideas can be an important part of professional development intended to help them examine and improve their own practices. Professional development can exploit TAs' internal conflicts between their learning goals and other considerations, such as, for example, considerations related to values of emotional and cognitive enjoyment. Professional development for TAs can directly address these conflicts, acquaint TAs with possible ways suggested in the educational literature to resolve similar conflicts, and allow the TAs to reflect on their practice in light of these new ideas and construct informed instructional choices.

In addition, the results of this study regarding TAs' conflict in the materialization of goals/considerations for choosing example solutions may extend well to faculty. As reported in a prior study [2], helping students develop an expert-like problem-solving approach is one of the prominent learning goals that the faculty expressed on the analysis of problem statements. However, their selection of problem features did not necessarily align with this goal because other conflicting values (e.g., the need for clarity in the presentation of problems and the concern about reducing the stress on students) drive them to include features that they believe hinder the goal. Although our current study with the TAs focuses on a different context (presenting example solutions to students), the conflict between instructional practices that support the goal of helping students develop expert-like problem-solving approaches and other goals/considerations appears to be common in both groups. Moreover, a pilot study with 6 faculty [3] has reported reasons why physics faculty may refrain from constructing example solutions that contain detailed explanation of expert thought processes. The reasons mirror those given by TAs and include: the value of concise representation of ideas, concerns regarding student engagement from emotional perspective, and the time required to construct a detailed solution. Further investigation of faculty beliefs would allow for a comparison between TA beliefs and faculty beliefs, which can provide valuable information to help researchers understand the pieces of instructor ideas related to the teaching of expert problem solving that develop naturally and those that need additional support to develop. If difficulties in materializing the goals coherently are confirmed with both TAs and faculty, the findings can also inform the design of interventions in order to help both groups come to terms with the conflict effectively.

Since our preliminary comparison suggests many faculty ideas are quite similar to those of the TAs, it is likely that not much learning spontaneously takes place without external intervention. The stage of TAs is a setting in which many TAs may be more amenable to learning from professional development than when they become faculty. Early exposure to new ideas about teaching and learning might put the TAs on a different learning trajectory, by allowing them to see aspects of teaching and learning that they would have otherwise been unaware of. According with the above, we reemphasize the importance of engaging TAs in effective professional development programs, in order to significantly improve the teaching practices in introductory physics classes.

## ACKNOWLEDGMENTS

We wish to thank the graduate teaching assistants at University of Pittsburgh, Department of Physics and Astronomy, who gave their valuable time to participate in this study. We wish to thank the PER group in the University of Minnesota, from which the solution artifacts used in this study were developed. We appreciate the support of the University of Pittsburgh, Department of Physics and Astronomy; Western Michigan, Department of Physics and Mallinson Institute for Science Education; and the Weizmann Institute of Science, Department of Science Teaching.

You are whirling a stone tied to the end of a string around in a vertical circle having a radius of 65 cm. You wish to whirl the stone fast enough so that when it is released at the point where the stone is moving directly upward it will rise to a maximum height of 23 meters above the lowest point in the circle. In order to do this, what force will you have to exert on the string when the stone passes through its lowest point one-quarter turn before release? Assume that by the time that you have gotten the stone going and it makes its final turn around the circle, you are holding the end of the string at a fixed position. Assume also that air resistance can be neglected. The stone weighs 18 N.

**FIG. 1. Problem used in the artifact comparison technique.**

## Instructor solution I

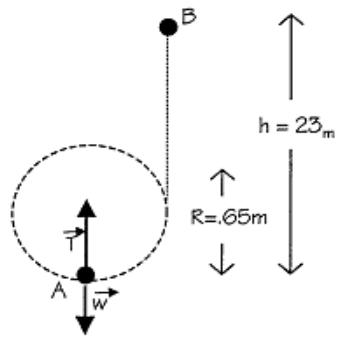

The tension does no work

Conservation of energy between point A and B

$Mv_A^2/2 = mgh$

$v_A^2 = 2gh$

At point A, Newton's 2nd Law gives us:

$\vec{T} - \vec{w} = m\vec{a}$

$T - w = mv_A^2/R$

$T = 18_N + 2 \cdot 18_N \cdot 23_m / .65_m =$ ☐ 1292N

**FIG. 2. Example solution artifact I.**

## Instructor solution II

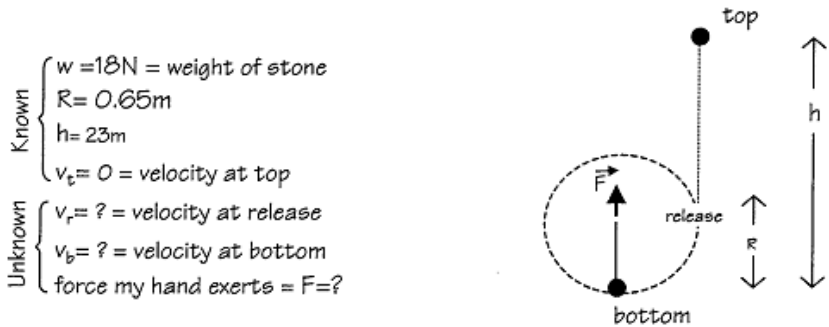

Known:
- w = 18N = weight of stone
- R = 0.65m
- h = 23m
- $v_t$ = 0 = velocity at top

Unknown:
- $v_r$ = ? = velocity at release
- $v_b$ = ? = velocity at bottom
- force my hand exerts = F = ?

Step 1) Find $v_r$ needed to reach h

$E_i = E_f$

$E_{release} = E_{top}$

$PE_{release} + KE_{release} = PE_{top} + KE_{top}$

$mgR + mv_r^2/2 = mgh + mv_t^2/2$

$v_r^2 = 2g(h - R)$

> Conservation of energy for the stone earth system, since no external forces.
> Note: you could also choose other systems.
> KE of earth estimated to be 0
> You could also use kinematics to find $v_r$.

Step 2) Find $v_b$ needed to have $v_r$ at release

$E_{bottom} = E_{release}$

$PE_{bottom} + KE_{bottom} = PE_{release} + KE_{release}$

$mg0 + mv_b^2/2 = mgR + mv_r^2/2$

Using $v_r$ from above:

$V_b = [2gh]^{1/2}$

> Conservation of energy for the stone earth system. Since T⊥v in circular path, T does no work.

Step 3) Find $T_b$, tension at bottom, needed for stone to have $v_b$ at bottom

$\sum \vec{F} = m\vec{a}$

$\sum F_R = ma_R$

$T_b - w = m v_b^2/R$

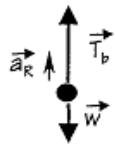

Free body diagram

> To relate the forces to velocity we can look at the radial component, and use $a_R = v^2/R$.

Using $v_b$ from above:

$T_b - w = 2 mgh/R$

$T_b = w + 2 w h/R = 18 + 2 \cdot 18 \cdot 23/.65 =$ $\boxed{1292N}$

$T_b$ equals F, the force my hand exerts, for a massless string

**FIG. 3. Example solution artifact II.**

## Instructor solution III

### Approach:

I need to find $F_m$, force exerted by me. I know the path, h (height at top) and $v_t$ (velocity at top)

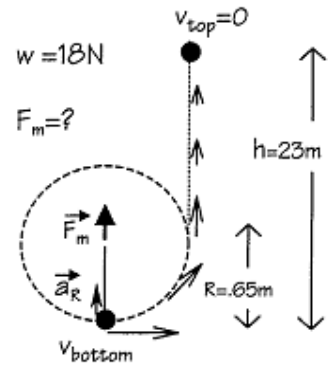

A) For a massless string $F_m = T_b$ ($T_b$-Tension at bottom)

B) I can relate $T_b$ to $v_b$ (velocity at bottom) using the radial component of $\sum \vec{F} = m\vec{a}$, and radial acceleration $a_R = v^2/R$, since stone is in circular path

C) I can relate $v_b$ to $v_t$ using either i) energy ii) Dynamics and kinematics
    ii) Messy since forces/accelerations change through the circular path
    i) I can apply work-energy theorem for stone. Path has 2 parts:
        first - circular, earth and rope interact with stone,
        second - vertical, earth interacts with stone
    In both parts the only force that does work is weight, since in first part hand is not moving $\Rightarrow \vec{T} \perp \vec{v} \Rightarrow \vec{T}$ does no work.

### Execution:

B) Relate $T_b$ to $v_b$

$\sum \vec{F} = m\vec{a}$

$\sum F_R = ma_R$

$T_b - w = m v_b^2/R$

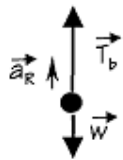

C) Relate $v_b$ to $v_t$

Work = $\Delta KE$

For constant force

$\vec{F} \cdot \vec{d} = KE_f - KE_i$

$F_y d_y = KE_{top} - KE_{bottom}$

$-w\,h = m v_t^2/2 - m v_b^2/2$

Substituting C) into B)

$T_b - w = 2 w h/R$

$F_m = T_b = w + 2 w h/R$

$= 18 + 2 \cdot 18 \cdot 23/.65$

$= \boxed{1292N}$

N=N m/m units O.K.

Large compared to weight, but stone needs to travel up large distance

Check limits: $T_b \uparrow$ as $R \downarrow$, for smaller circle I'll need bigger force, reasonable

**FIG. 4. Example solution artifact III. (The equation in gray was not shown on the solution provided to the TAs due to a problem that occurred in the copying process.)**

In what situations do you believe students should be provided with examples of solved problems? (e.g., during lecture, after homework or a test, etc.) What is the purpose you see for providing solved examples in these different situations? How would you like your students to use the solved examples you give them in these different situations? Why? What do you think most of them actually do?

**FIG. 5. Open-ended questions.**

| Attached are several instructor solutions for the problem you solved that were designed to be posted or distributed to students. They are based on actual instructor solutions. Take a look at each of these instructor solutions and describe the prominent features of those solutions. Which features of these solutions would you like to include in solutions you are writing for your students? Please explain your reasons | | | | | | | |
|---|---|---|---|---|---|---|---|
| Solution features | Rank the solutions based on which solution has more of this feature. (You could also Mark + for the solutions in which this feature exists.) | | | Rank the solutions based on your **preference** for this feature (A - for the one you like the most in how it represents this feature to C-for the one you like the least) | | | Why do you like/ not like this feature? |
| | Sol. I | Sol. II | Sol. III | Sol. I | Sol. II | Sol. III | |
| | | | | | | | |

**FIG. 6. Pre-discussion worksheet.**

| Write down the features' numbers that you originally noticed (using attached feature list (Table II). You might have termed them somewhat differently.) For each of your features, please: Write down how you originally named this feature. Describe how and why, if at all, your preference towards it changed following the class discussion. | | | | | |
|---|---|---|---|---|---|
| Feature number | Your original feature name | Rate the solutions based on your current preference for this feature | | | In case your preference towards it changed following the class discussion, elaborate your final preferences: Why do you like or dislike this feature? |
| | | Sol. I | Sol.II | Sol. III | |
| | | | | | |

**FIG. 7. Post-discussion worksheet.**

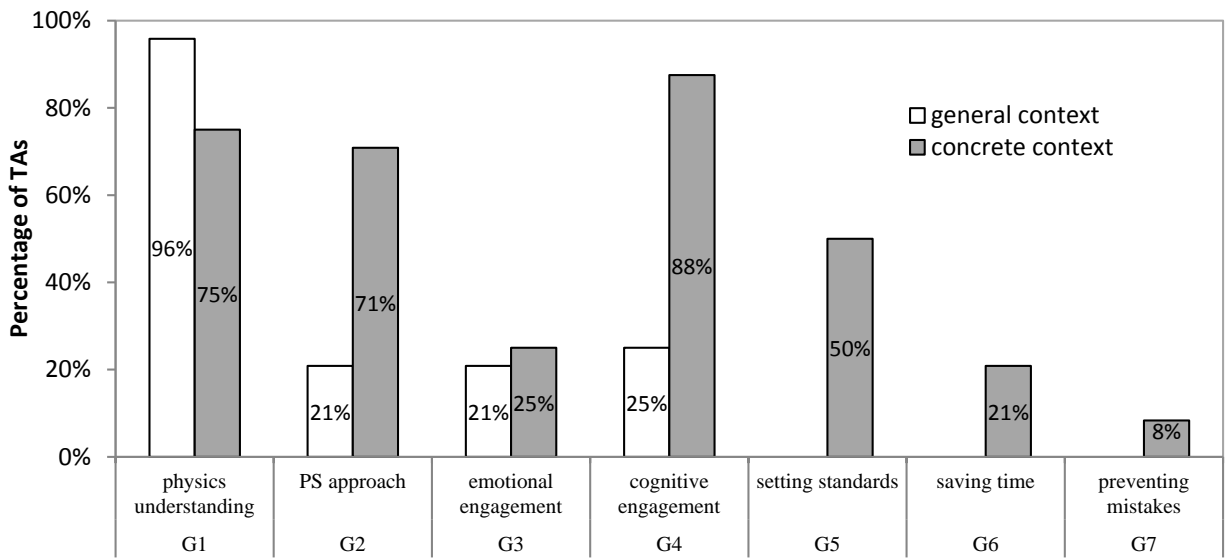

**FIG. 8. Percentage of TAs who mentioned each goal/consideration in the general and concrete contexts, respectively.**

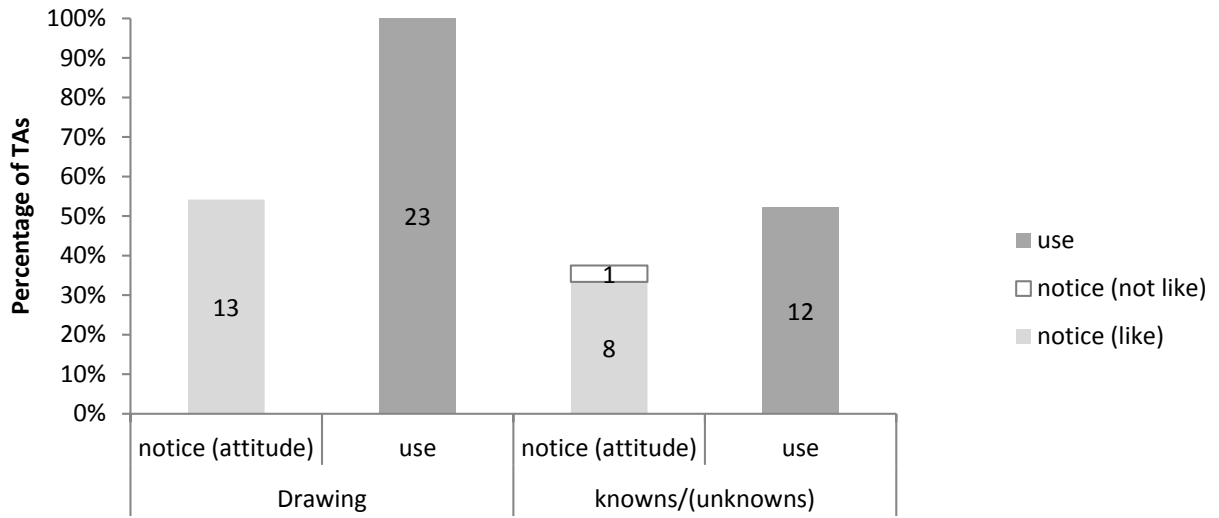

**FIG. 9.** Percentages of TAs who noticed (total N = 24) vs. used (total N = 23) features related to "initial problem analysis". For TAs who noticed the features, their attitudes are also presented. The like/not like rating for each feature was determined by examining the reasons that the TA wrote on the worksheet for why they like or do not like the feature and comparing whether the solution that each TA ranked as highest based on his/her preference for the feature matches the solution that he or she believed contain the most of this feature. The number of TAs in each case is also labeled.

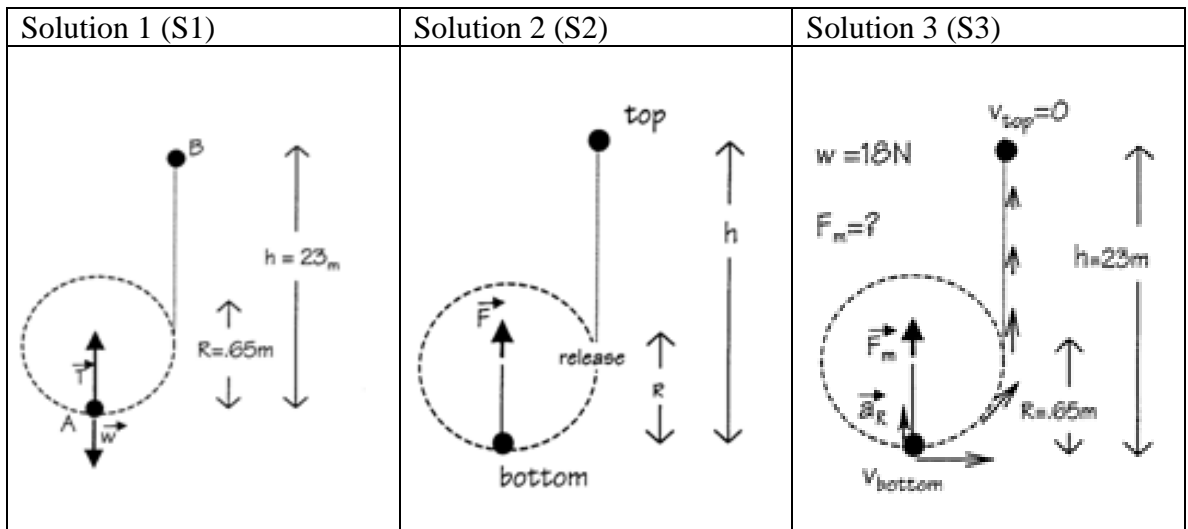

**FIG. 10.** Diagram used in each of the 3 example solution artifacts.

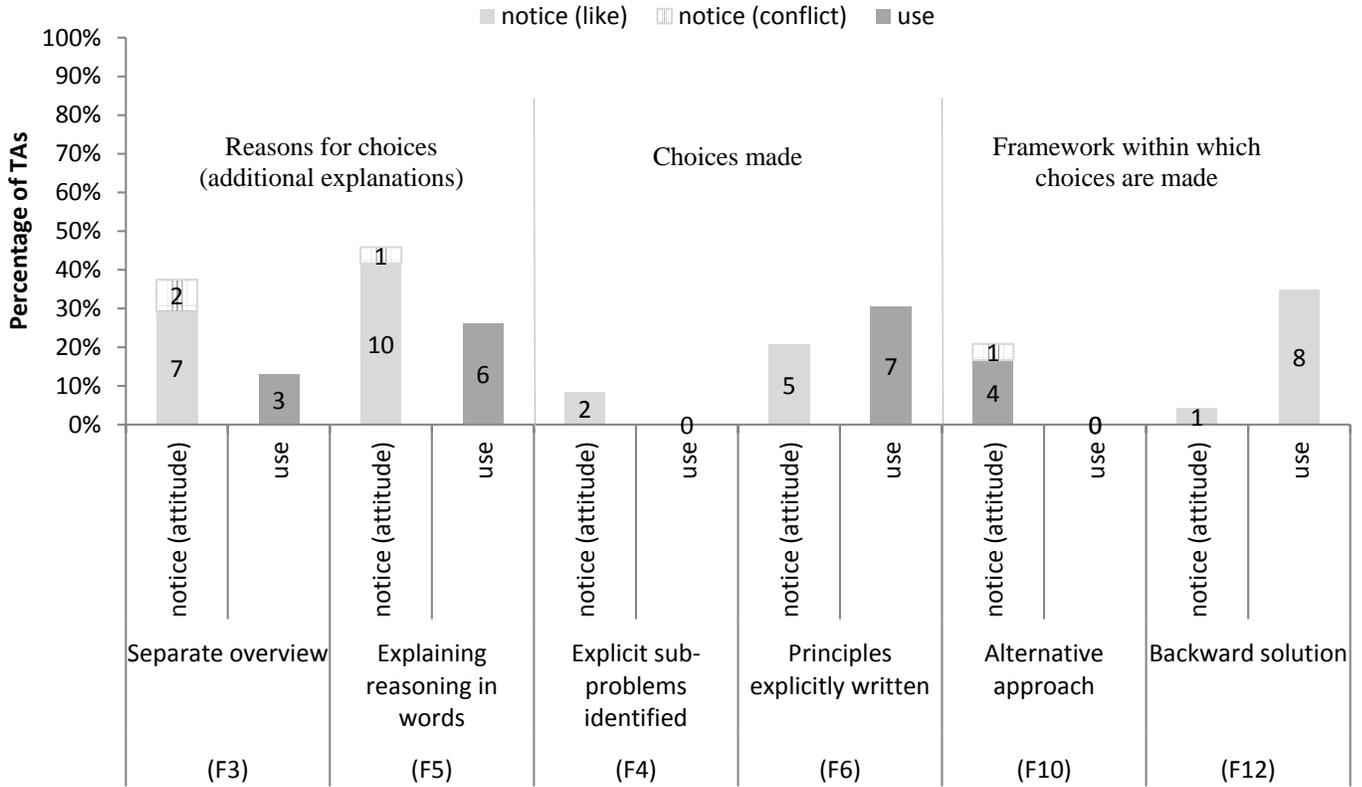

**FIG. 11.** Percentage of TAs who noticed (N=24) vs. used (N=23) features related to "solution construction". For TAs who noticed the features, their attitudes are also presented. The number of TAs in each case is also labeled.

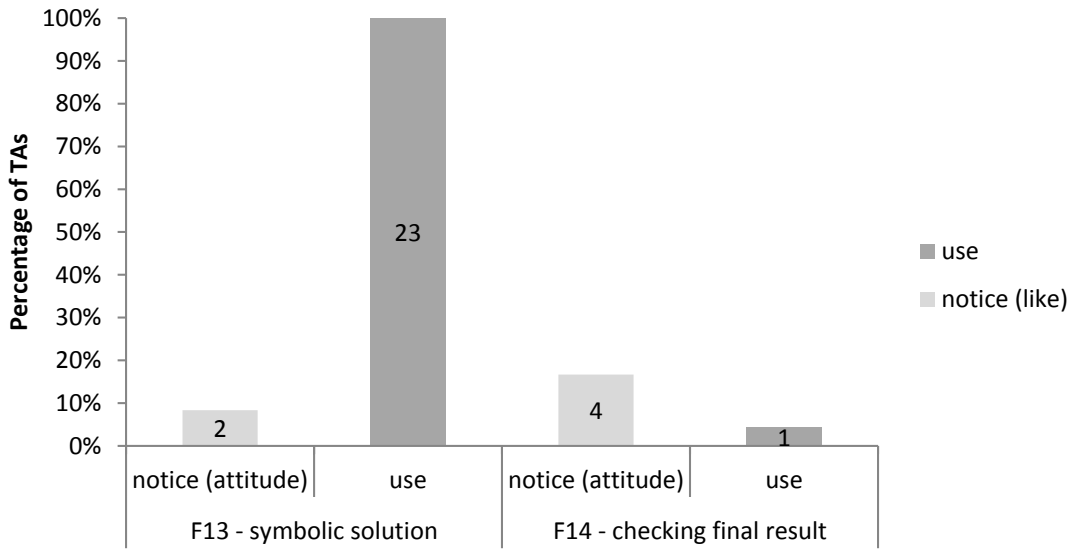

**FIG. 12.** Percentage of TAs who noticed (N=24) vs. used (N=23) features related to "checking of solution". For TAs who noticed the features, their attitudes are also presented. The number of TAs in each case is also labeled.

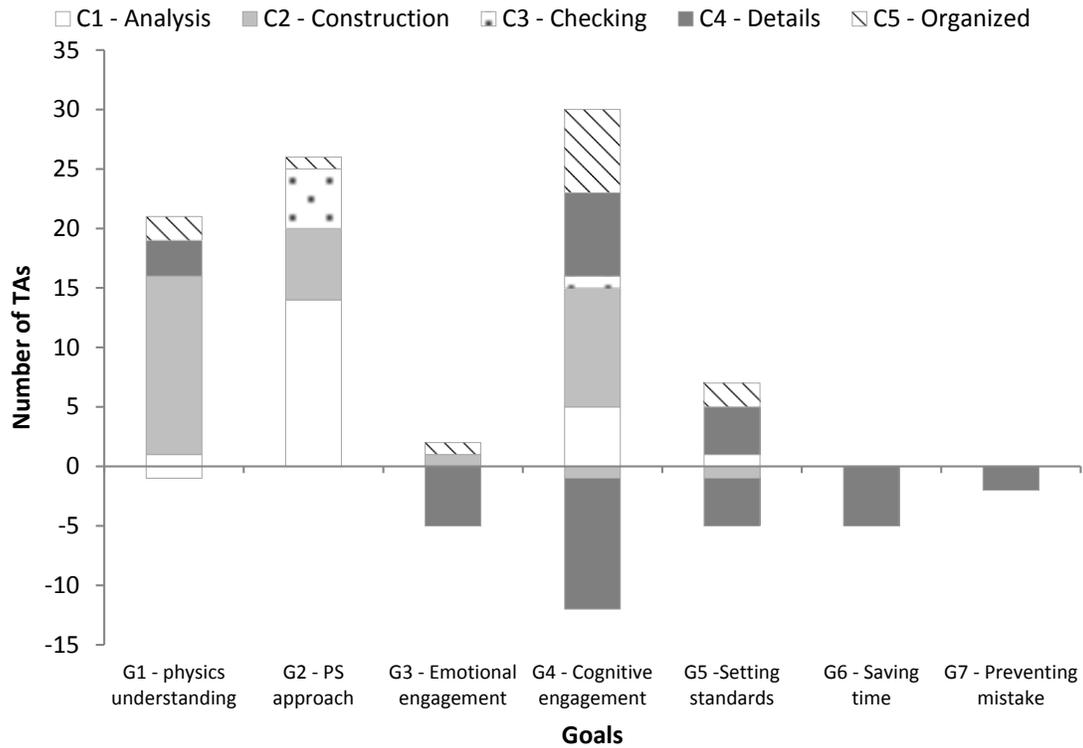

FIG. 13. Number of TAs who mentioned features as supportive or contradictive to the goals (Gs). To get a somewhat more global picture, the 14 features are compressed into 5 clusters (Cs) as described in Table 2.

**Table 1. GAIQ sequence of activities.**

| Time | Activity |
|---|---|
| Pre | Individually, TAs solve target problem (FIG. 1) and answered open-ended questions as well as **questions** in pre-discussion worksheet (FIG. 6) that are related to the 3 example solutions. |
| Lesson | In groups of 3, TAs answered the **same questions** in group worksheets, then a whole class discussion took place where groups share their work. |
| Post | Individually, TAs answered **same questions** in post-discussion worksheet (FIG. 7). They were also asked to match the features they identified on the pre-discussion worksheet to a list of pre-defined features presented in Table 2. (TAs were only given the descriptions of each feature and not the information about clusters). |

**Table 2. Pre-defined feature list (from pilot study). Features related to the key stages in a prescriptive problem solving model are grouped into clusters C1 to C3. Features related to the communication of the solution are grouped into clusters C4 and C5.**

| Feature | Description | Cluster |
|---|---|---|
| F1 | Provides a schematic visualization of the problem (a diagram) | C1- Initial problem analysis |
| F2 | Provides a list of knowns/unknowns | C1 - Initial problem analysis |
| F3 | Provides a "separate" overview of how the problem will be tackled (Explains premise and concepts -- big picture -- prior to presenting solution details) | C2- Solution construction |
| F4 | Explicit sub-problems are identified (Explicitly identifies intermediate variables and procedures to solve for them) | C2- Solution construction |
| F5 | Reasoning is explained in explicit words (Description/justification of why principles and/or subproblems are appropriate/useful in this situation) | C2- Solution construction |
| F6 | The principles/concepts used are explicitly written using words and/or basic mathematical representations (e.g., F=ma or Newton's $2^{nd}$ Law) | C2- Solution construction |
| F7 | Thorough derivation (Detailed/verbose vs. Concise/short/simplified/ skips lots of derivation) | C4-Extended details |
| F8 | Long physical length (Long/verbose vs. Short/concise vs. Balanced/not too long, not too short) | C4-Extended details |
| F9 | Includes details that are not necessary for explaining the problem solution (The solution is technically correct and complete without these 'unnecessary' details) | C4-Extended details |
| F10 | Provides alternative approach | C2- Solution construction |
| F11 | Solution is presented in an organized and clear manner | C5-Organization and clarity |
| F12 | Direction for the progress of the solution progress: Backward vs. forward | C2- Solution construction |
| F13 | Symbolic solution (Numbers are plugged-in only at the end) | C3-Checking of solution |
| F14 | Provides a check of the final result (e.g. if the unit is correct, or if the answer makes sense by examining the limits) | C3-Checking of solution |

**Table 3** The different goals/considerations that the TAs expressed when discussing the use of example solutions (corresponding to the general context) and specific design features that they would/would not use in their solutions (corresponding to the concrete context). The context(s) in which each goal/consideration is mentioned are listed. The meanings of Goal 3 and Goal 4 differ slightly in different contexts and therefore a separate description is given for each context.

| Goals | Context | Example solutions should… |
|---|---|---|
| G1: Physics understanding | Both | …help students construct content specific physics understanding. |
| G2: PS approach | Both | …help students develop an expert-like problem solving (PS) approach. |
| G3: Emotional engagement | General | …be used to motivate students or to prevent student frustration. |
| | Concrete | …be designed in a way to maintain students' interest. |
| G4: Cognitive engagement | General | …be presented at a proper time to engage students in cognitive processing. |
| | Concrete | …be communicated in a manner that students can follow. |
| G5: Setting standards | Concrete | …be designed in a way that demonstrate the standard of an adequate solution. |
| G6: Saving time | Concrete | …be designed to save TA and student time. |
| G7: Preventing mistakes | Concrete | …help students avoid losing points on tests. |

**Table 4.** Comparison of the number of TAs who (1) noticed either F3 and/or F5 vs. F7 and/or F8 (either one of them) and (2) expressed positive (Like), negative (Don't Like), or mixed preference for the feature(s) in the pre-discussion worksheet

| | Former undergraduate education: USA (N=11) | | Former undergraduate education: Other (N=13) | |
|---|---|---|---|---|
| | Notice | Preference | Notice | Preference |
| F3 or F5 | 9/11 (82%) | Like:9/9 (100%) | 7/13 (54%) | Like: 6/7 (86%)<br>Mixed: 1/7 (14%) (*) |
| F7 or F8 | 8/11 (73%) | Like: 4/8 (50%)<br>Don't Like : 4/8 (50%) | 12/13 (92%) | Like: 1/12 (8.3%)<br>Don't Like: 9/12 (75%)<br>Mixed: 2/12 (17%) (**) |

(*): This TA noticed F5, which he originally named as "marginal notes" and in general valued it. He explained that this feature "*give notes for some procedures*". However, he also added a comment saying that "*but it's not good for too many notes*".

(**): One TA noticed F8 and indicated that there are pros and cons for a concise solution. He explained that a concise solution "*saves time, but could also cause confusion*". Overall speaking, this TA liked solution 1 (the concise solution) the best. The other TA expressed a somewhat conflicting preference between F7 and F8. He valued F7, which he originally named as "sufficient details", but preferred a brief demonstration when discussing F8.

[i] The training course aimed to guide TAs to contemplate issues related to the teaching of physics and explore possible strategies to improve their teaching.